# QVRF: A QUANTIZATION-ERROR-AWARE VARIABLE RATE FRAMEWORK FOR LEARNED IMAGE COMPRESSION

*Kedeng Tong*[1*], *Yaojun Wu*[2], *Yue Li*[2], *Kai Zhang*[2], *Li Zhang*[2], *Xin Jin*[1]

[1]Shenzhen International Graduate School, Tsinghua University, Shenzhen, 518055 China
[2]Multimedia Lab, Bytedance Inc., San Diego CA., 92122 USA

## ABSTRACT

Learned image compression has exhibited promising compression performance, but variable bitrates over a wide range remain a challenge. State-of-the-art variable rate methods compromise the loss of model performance and require numerous additional parameters. In this paper, we present a Quantization-error-aware Variable Rate Framework (QVRF) that utilizes a univariate quantization regulator *a* to achieve wide-range variable rates within a single model. Specifically, QVRF defines a quantization regulator vector coupled with predefined Lagrange multipliers to control quantization error of all latent representation for discrete variable rates. Additionally, the reparameterization method makes QVRF compatible with a round quantizer. Exhaustive experiments demonstrate that existing fixed-rate VAE-based methods equipped with QVRF can achieve wide-range continuous variable rates within a single model without significant performance degradation. Furthermore, QVRF outperforms contemporary variable-rate methods in rate-distortion performance with minimal additional parameters. The code is available at https://github.com/bytedance/QRAF.

## 1. INTRODUCTION

Lately, the thriving image compression techniques [1-11] that utilize the Variational-AutoEncoder (VAE) framework have demonstrated promising and superior compression performance when compared to hybrid codecs like JPEG [12], JPEG2000 [13], HEVC [14], and VVC [15]. These large-receptive-field learned compression methods are trained on large datasets to minimize the rate-distortion (RD) cost $R + \lambda D$ and to optimize network parameters globally for various contents. Here, $\lambda$ is the Lagrange multiplier to control the RD trade-off. Those methods overcome the disadvantage of region-limited prediction and module-separately optimization presented in hybrid codecs. Thus, learned image compression is considered to be the future of image compression and JPEGAI is committed to creating a learning-based image coding standard [16].

However, achieving variable bitrate over a wide range still poses a significant challenge for learned image compression. Most learned methods cover a wide range of bitrates by training multiple models that correspond to multiple values of $\lambda$ and each model is fixed-rate only. The use of multiple fixed-rate models results in a considerable cost of training and memory requirements, and they are unable to adjust bitrate flexibly like hybrid codecs, which can achieve a wide range of variable rates and rate adaption in one codec without performance degradation (e.g., 51 levels for HEVC) [17-19].

To address this issue, certain techniques employ a single model to achieve variable rates. The conditional autoencoder schemes [20-24] incorporate subnetworks, which take the prior information of $\lambda$ as input, into convolutional layers to attain variable rates. However, these schemes have considerably increased the computational complexity and model size. Choi *et al.* [24] have suggested adjusting the interval size of uniform noise in universal quantization to extend the fixed-rate model to a narrow range of continuous variable rates (e.g. ±0.1 bpp) without significant performance degradation. The amplitude modulation paradigm [25-28] defines scaling coefficients or subnetworks to modify the amplitude of the latent, thereby indirectly controlling the quantization error and attaining variable bitrates. Exponent interpolation [26] of the scaling coefficients contributes to the attainment of continuous variable rates. Nevertheless, the paradigm has an impact on the distribution of probability modeling and demonstrates compression performance degradation in wide-range variable bitrates.

In this paper, we propose a Quantization-error-aware Variable Rate Framework (QVRF) which utilizes a univariate quantization regulator *a* to control overall quantization error of latent representation to achieve wide-range variable rates within a single model. Specifically, the QVRF defines a quantization regulator vector coupled with a predefined Lagrange multiplier set for discrete variable rates. Additionally, we propose a reparameterization method to enable QVRF to function compatibly with a round quantizer. We find that the *a* and the square root of $\lambda$ are linear and changing the value of *a* in the quantization and

---

entropy process is sufficient to achieve continuous variable rates without performance degradation in QVRF. To demonstrate its generality, we integrate QVRF into three learned image compression methods [8-10]. The VAE-based models equipped with QVRF achieve wide-range continuous variable rates within a single model. Notably, QVRF outperforms State-of-the-art (SOTA) variable rate methods in terms of RD performance, additional parameters and computational complexity.

## 2. PROPOSED METHOD

In this section, we first review the formulation of VAE-based compression. Then, we introduce the proposed QVRF and reparameterization method.

### 2.1. Formulation of VAE-based model

The optimization problem of the VAE-based compression can be formulated as a Lagrange optimization process:

$$\mathcal{L} = \mathcal{R}(\hat{y}) + \mathcal{R}(\hat{z}) + \lambda \cdot \mathcal{D}(x, \hat{x})$$
$$= \mathbb{E}[-\log_2(p_{\hat{y}|\hat{z}}(\hat{y}|\hat{z}))] + \mathbb{E}[-\log_2(p_{\hat{z}}(\hat{z}))]$$
$$+ \lambda \cdot \mathcal{D}(x, \hat{x} | \hat{y}) \quad , \quad (1)$$

where $\mathcal{R}(\cdot)$, $\mathcal{D}(\cdot)$, $p_{\hat{z}}$ and $p_{\hat{y}|\hat{z}}(\hat{y}|\hat{z})$ are compression bitrate, the distortion between the raw image $x$ and the reconstructed image $\hat{x}$, the probability mass function of quantized size information $\hat{z}$, and the Gaussian entropy estimation of quantized latent representation $\hat{y}$ on condition of $\hat{z}$, respectively. Due to the round quantization for lossless arithmetic coding, the entropy model of quantized latent $\hat{y}$ is formulated as:

$$p_{\hat{y}|\hat{z}}(\hat{y}|\hat{z}) = \prod_i p_{\hat{y}|\hat{z}}(\hat{y}_i|\hat{z})$$
$$p_{\hat{y}|\hat{z}}(\hat{y}_i|\hat{z}) = \mathcal{N}(u_i, \sigma_i) * \mathcal{U}(-\tfrac{1}{2}, \tfrac{1}{2})(\hat{y}_i) \quad , \quad (2)$$

where $\hat{y}_i$, $u_i$, and $\sigma_i$ represent the $i$-th element of $\hat{y}$, the mean and the scale of the estimated Gaussian distribution, respectively. The optimization problem under the fixed $\lambda$ in Eq. 1 is employed for the fixed-rate paradigm.

### 2.2. Quantization-error-aware rate adaption framework

In contrast to interval adjustment of universal quantization under a fixed $\lambda$ [24], we proposed to couples $\lambda$ and $a$, a univariate quantization regulator, to control the quantization error of overall latent representation and achieve wide-range variable rates. The framework of the proposed QVRF is shown in Fig. 1, where $Q$ and $AE/AD$ denote the quantization and arithmetic entropy coding, respectively. As depicted in the figure, the QVRF uses the univariate $a$ in quantization and entropy coding process to control the overall quantization error of latent representation $y$ for variable rates. QVRF keeps the input of the probability modeling network unchanged, maintaining the Gaussian entropy estimation of $y$ nearly fixed. The quantization regulator $a$ serves as a mechanism to modify quantization bin size. For each element of latent representation $y_i$ generated by the encoder in QVRF, the mass cumulative function in Eq. 2 is formulated as:

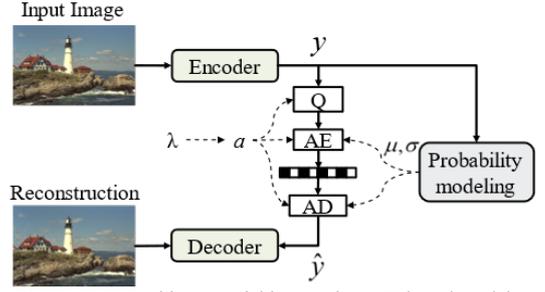

**Fig. 1**. QVRF. QVRF achieves variable rates in VAE-based models using a univariate quantization regulator $a$ to control the quantization error of all $y$ in the quantization and entropy coding process while maintaining a nearly fixed probability modeling.

$$p_{\hat{y}|\hat{z}}(\hat{y}_i|\hat{z}, a) = \int_{y_i - \frac{1}{2a}}^{y_i + \frac{1}{2a}} \frac{1}{\sqrt{2\pi}\sigma_i} e^{-\frac{(y_i - u_i)^2}{2\sigma_i^2}} dy_i$$
$$= \mathcal{N}(u_i, \sigma_i) * \mathcal{U}(-\tfrac{1}{2a}, \tfrac{1}{2a})(\hat{y}_i) \quad . \quad (3)$$

And the quantization error $QE$ between $y$ and $\hat{y}$ in Eq. 3 is:

$$QE = y - \hat{y} = \mathcal{U}(-\tfrac{1}{2a}, \tfrac{1}{2a}) . \quad (4)$$

The preceding analysis confirms that adjusting the value of $a$ equal to alter the overall quantization error between $y$ and $\hat{y}$, and impact the required bits for storage and transmission. We propose a coupling between $\lambda$ and $a$, establishing a one-to-one correspondence between the two. The prior works [25, 26] demonstrate that using multiple values of $\lambda$ in training is necessary for discrete variable rates. Thus, a quantization regulator vector $A \in R^n$, made of $\{a_1, ..., a_n\}$, is defined and assembled from in the training process, mapping to the predefined Lagrange multiplier set $\Lambda$ containing $n$ elements. Moreover, the RD optimization function in Eq. 1 is altered into:

$$\mathcal{L} = \sum_{\lambda \in \Lambda} \mathcal{R}(\hat{y}) + \mathcal{R}(\hat{z}) + \lambda \cdot \mathcal{D}(x, \hat{x})$$
$$= \sum_{j=1}^{n} \mathbb{E}[-\log_2(p_{\hat{y}|\hat{z}}(\hat{y}|\hat{z}, a_j))] + \mathbb{E}[-\log_2(p_{\hat{z}}(\hat{z}))] . \quad (5)$$
$$+ \lambda_j \cdot \mathcal{D}(x, \hat{x} | \hat{y})$$

### 2.3. Reparameterization method

Reparameterization method is proposed to make QVRF compatible with existing learned compression methods [8-10] that utilize a round quantizer for quantization and arithmetical coding. Reparameterization method involves replacing Eq. 3 with a scaling operation on $y_i$, $u_i$, and $\sigma_i$:

$$p_{\hat{y}|\hat{z}}(\hat{y}_i|\hat{z}) = \int_{ay_i - \frac{1}{2}}^{ay_i + \frac{1}{2}} \frac{1}{\sqrt{2\pi}(a\sigma_i)} e^{-\frac{(ay_i - au_i)^2}{2(a\sigma_i)^2}} day_i$$
$$= \mathcal{N}(au_i, a\sigma_i) * \mathcal{U}(-\tfrac{1}{2}, \tfrac{1}{2})(a\hat{y}_i) \quad . \quad (6)$$

Thus, scaled $y_i$ can be quantized by the round quantizer while containing quantization error of $1/a$, and the scaled quantized values are used for entropy coding. Note that the scaled quantized value for entropy coding ought to be inversed-scaled by a strictly reciprocal principle to obtain accurate $\hat{y}_i$.

## 3. EXPERIMENTAL RESULTS

To demonstrate the effectiveness of our proposed algorithm, experiments and comparisons are conducted.

### 3.1. Experimental settings

Following the previous work [4], a subset of 8,000 images from ImageNet training set [29] and 584 images from CLIC2020 training set are [30] selected for our training data. The Mean Square Error (MSE) serves as the distortion in Eq. 5. We choose three representative VAE-based methods including the basic VAE-based method [10], autoregressive VAE-based model [9], and advanced network with attention module [8], as our baselines. We incorporate QVRF with each of them. To cover from the low bitrates to high bitrates, we defined the predefined Lagrange multiplier set $\Lambda$ as (0.0018, 0.0035, 0.0067, 0.0130, 0.0250, 0.0483, 0.0932, 0.1800). Empirically, the first value in $\Lambda$ is set to the reference $\lambda_{ref}$, and $A$ is initialized as $\sqrt{\lambda/\lambda_{ref}}$ with $a_1$ set to constant 1. The training process uses the Adam optimizer, with a batch size of 8 and the learning rate initially set to 0.0001. We use staged training strategies, where the network parameters are optimized on $\lambda$=0.18 for the first 2000 epochs. Then, $A$ are optimized jointly with noise approximation for 500 epochs and straight-through estimation [5] for another 500 epochs. The bitrate, measured by bit per pixels (bpp), is actual bits of entropy coding in the RD curves. Peak Signal-to-Noise Ratio (PSNR) and Multi-Scale-Structural Similarity Index measure (MS-SSIM) [16] are objective distortion criteria for image distortion.

We conduct testing of our proposed algorithm on several datasets including the Kodak dataset with 24 images [31], the CLIC validation dataset with 41 high-resolution images [30], the JEPGAI test dataset with 16 high-quality images [16]. We include fixed-rate models of Ballé et al. [10] and Minnen et al. [9] from [32]. We re-train 8 fixed-rate models of Cheng et al. [8], where convolution channels $N$ and $M$ are both 192, on the same training dataset for comparison, To ensure a fair comparison, we implement three SOTA variable rate methods [21, 26, 27] on the baseline models following the original paper settings.

### 3.2. Discrete and continuous variable bitrates

As shown in Fig.2, QVRF can achieve both discrete and continuous variable bitrate (DVR/CVR) within a single model with negligible performance degradation compared to eight retrained fixed-rate models based on Cheng's method [8] serving as the benchmark. Quantization regulator vector $A$ is jointly optimized with the network parameters by minimizing the loss function Eq. 5. Consequently, the trained model can accomplish discrete variable rates by selecting an appropriate value of $a$ from $A$ for discrete variable bitrates. Owing to the compatibility of wide-range values of $a$ and network parameters, altering the value of $a$, shared for the encoding and decoding process, within the expanded range of the minimum value of $A$ to the maximum value of $A$, can achieve continuous variable bitrates, as presented in Fig. 2.

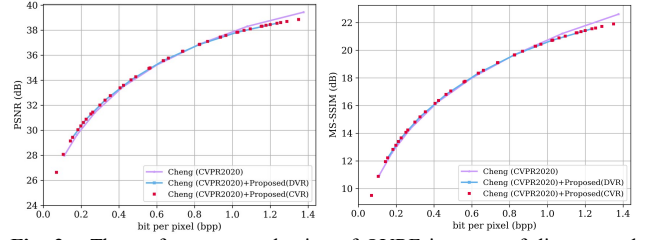

**Fig. 2**. The performance evaluation of QVRF in terms of discrete and continuous variable rates is conducted using 8 fixed-rate models of Cheng et al. [8] as the benchmark.

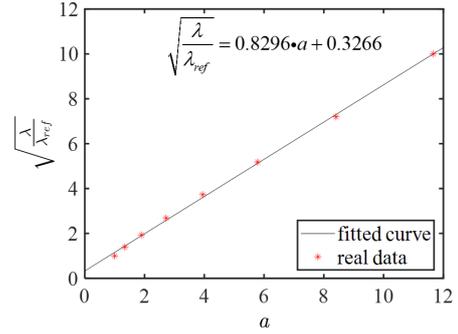

**Fig. 3**. The linear correlation between univariate quantization regulator $a$ and the squared root of Lagrange multiplier $\lambda$. The real data is obtained from the retrained model of QVRF and Cheng et al. [8]. The value of $a$ corresponding to the reference Lagrange multiplier $\lambda_{ref}$ is set to the constant 1.

As depicted in Fig. 3, a linear relationship between $a$ and the squared root of $\lambda$ can be observed in a trained model with QVRF and Cheng et al. [8]. The analysis is based on the real data derived from the trained $A$ and $\Lambda$. $a_1$ corresponds to $\lambda_{ref}$ is constant 1. The linear function that fits the relationship between $a$ and squared root of $\lambda$ can be expressed as follows:

$$\sqrt{\frac{\lambda}{\lambda_{ref}}} = 0.8296 \cdot a + 0.3266 \qquad (7)$$

It reveals explicit linear relationship of $a$ and the squared root of $\lambda$ in the QVRF for variable rates and may contribute to rate adaption.

### 3.3. Comparison of variable-rate methods

As demonstrated in Fig. 4, QVRF outperforms existing variable bitrate techniques [21, 26, 27] for all three baseline models. Theis et al. [27] exhibits obvious performance degradation at both low and high bitrates. The RD performance of Song et al. [21] on Ballé et al. [10] produces superior outcomes compared to fixed-rate models at low bitrates, but presents inferior performance at midrates compared to QVRF. Cui et al. [26] attains comparable compression performance to fixed-rate models at low and middle bitrates but exhibits more discernible performance degradation than QVRF.

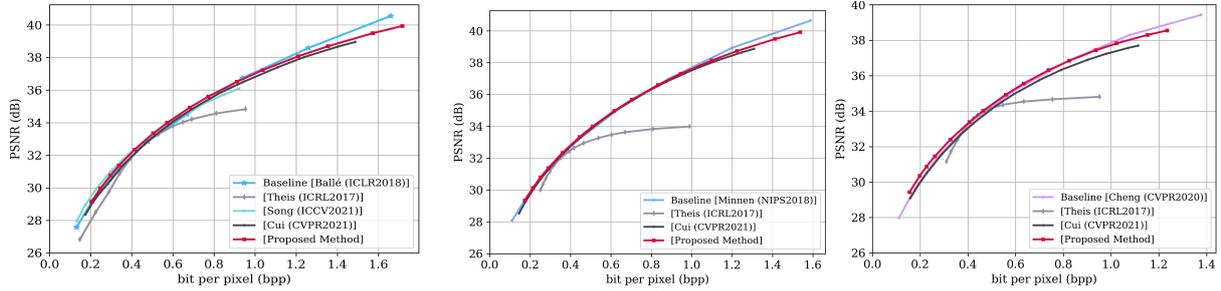

**Fig. 4**. Results of variable rate methods [22, 27, 28] and QVRF with three baselines on Kodak dataset. The baseline models are Ballé *et al.* [10], Minnen *et al.* [9], and Cheng *et al.* [8] from left to right.

We evaluate the performance of Cui *et al.* [26] and QVRF on BD-Rates and multi-distortion trade-off to show the superiority of QVRF. Table 1 displays the BD-Rates results of QVRF and Cui *et al.* [26] compared to fixed-rate methods [8, 10] on the Kodak dataset. BD-Rates are calculated with four RD points chosen from low, medium and high bitrates. Cui *et al.* [26] exhibits an increase of 2.58%, 1.29%, 4.87% in bitrates compare to fixed-rate models Ballé *et al.* [10], Minnen *et al.* [9], and Cheng *et al.* [8], respectively. Due to the improvements at low bitrate and slight performance degradation at high bitrates, QRVF achieves a bitrate reduction of 0.25%, -0.08%, 1.21% on three base models, respectively. For multi-distortion trade-off, we test the QVRF and Cui *et al.* [26] on CLIC validation [30] and JEPGAI test dataset [16] using Minnen *et al*. [9] as the baseline model. As shown in Fig. 5 and Fig. 6, Cui *et al.* [26] displays a narrower variable bitrates without performance degradation than QVRF in terms of PSNR and MS-SSIM compared to fixed-rate models.

Aside from compression performance, additional parameters and complexity are also important metrics for determining the feasibility of variable rate methods. We compare the additional parameters *Para.* and computational complexity FLOPs of variable rate methods [21, 26, 27] and QVRF with *n* set to 8. The results of this comparison, expressed as percentages (%) relative to a single fixed-rate model, are presented in Table 2. Notably, QVRF utilizes minimal additional parameters, and is nearly 200 times smaller in this regard than Theis *et al.* [27] and Cui *et al.* [26] on additional parameters. Furthermore, QVRF uses additional FLOPs percentages similar to those of Theis *et al.* [27] and Cui *et al.* [26], but is nearly 300,000 times smaller than Song *et al.* [21].

## 4. CONCLUSIONS

By incorporating a univariate quantization regulator *a* into the quantization and entropy coding process, the proposed QVRF enhances the existing VAE-based compression methods to achieve continuous variable rates within a single model. The reparameterization method makes QVRF compatible with the round quantizer. We demonstrated that square root of Lagrange multiplier and *a* are linear and altering the value of *a* can achieve continuous variable rates.

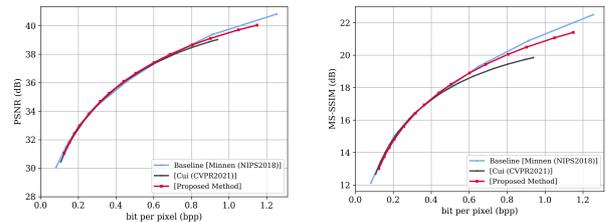

**Fig. 5**. Performance evaluation on CLIC validation dataset.

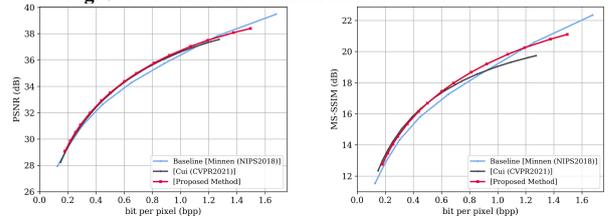

**Fig. 6**. Performance evaluation on JPEGAI test dataset.

**Table 1.** Average BD-Rate results of Cui *et al.* [26] and QVRF vs. fix-rate models [8, 10] on the Kodak dataset.

| Variable method | Fixed-rate model | | |
|---|---|---|---|
| | Ballé *et al.* [10] | Minnen *et al.* [9] | Cheng *et al.* [8] |
| Fixed-rate model + Cui *et al.* [26] | 2.58% | 1.29% | 4.87% |
| Fixed-rate model + QVRF | **-0.25%** | **0.08%** | **-1.21%** |

**Table 2.** The percentages (%) of additional parameters and complexity of variable rate methods [22, 27, 28] and QVRF compared to a single fix-rate model.

| Basic Model | Theis *et al.* [27] | | Song *et al.* [21] | | Cui *et al.* [26] | | QVRF | |
|---|---|---|---|---|---|---|---|---|
| | *Para.* | FLOPs | *Para.* | FLOPs | *Para.* | FLOPs | *Para.* | FLOPs |
| Ballé *et al.* [10] | 0.02168 | 0.00060 | 251.77958 | 268.37507 | 0.06932 | 0.00062 | **0.00007** | 0.00089 |
| Cheng *et al.* [8] | 0.00691 | 0.00015 | - | - | 0.02073 | 0.00016 | **0.00003** | 0.00022 |

Experimental results demonstrate that QVRF facilitates existing learned methods to achieve wide-range continuous variable rates within a single model while remaining comparable performance to the fixed-rate models. Additionally, QVRF outperforms the current variable rate methods in terms of RD performance, additional parameters and complexity.

# APPENDIX

**Diagram of variable rates in QVRF and reparameterization method**

As depicted in Fig. A1, using a value of *a*>1 leads to a reduction in the probability of quantized values, necessitating an increase in the number of bits used in entropy coding.

**Performance improvement of QVRF at high bitrate**

Merely relying on a larger pre-defined set of Lagrange multipliers can substantially improve the performance of QVRF at a high bitrate. As illustrated in Fig. A2, when the proposed combined model utilizes a pre-defined set of Lagrange multipliers (0.0018, 0.0035, 0.0067, 0.0130, 0.025, 0.0483, 0.0932, 0.18, 0.36, 0.72, 1.44), it exhibits comparable compression performance to the 8 fixed-rate models with lambda in the range of (0.0018, 0.0035, 0.0067, 0.0130, 0.0250, 0.0483, 0.0932, 0.1800), across all bitrates.

In Fact, the combined model achieves a 0.4% bitrate reduction relative to the fixed-rate models.

**Bit consumption analysis**

Based on the results of the bitrate consumption and compression performance presented in Table R1, it can be found that QVRF only has an impact on the compression of the latent representation, which requires a fixed number of bits for the side information.

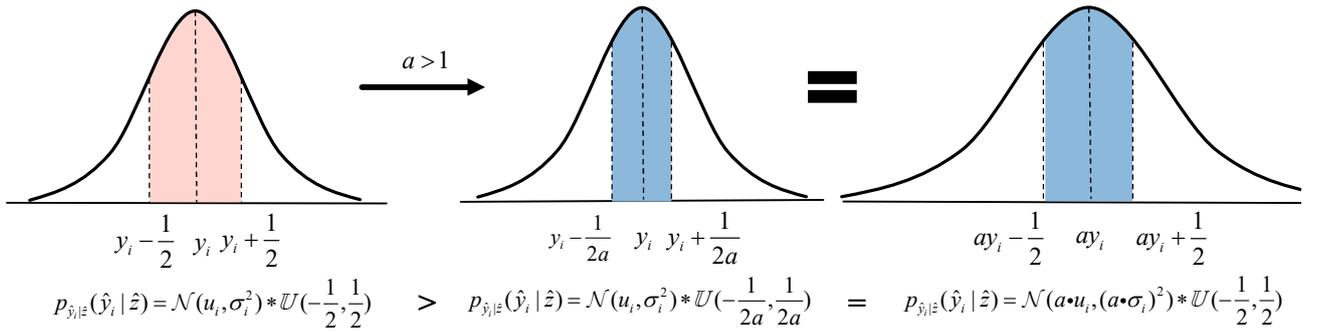

Fig. A1 The entropy coding requires more bits for quantized value with a larger *a*.

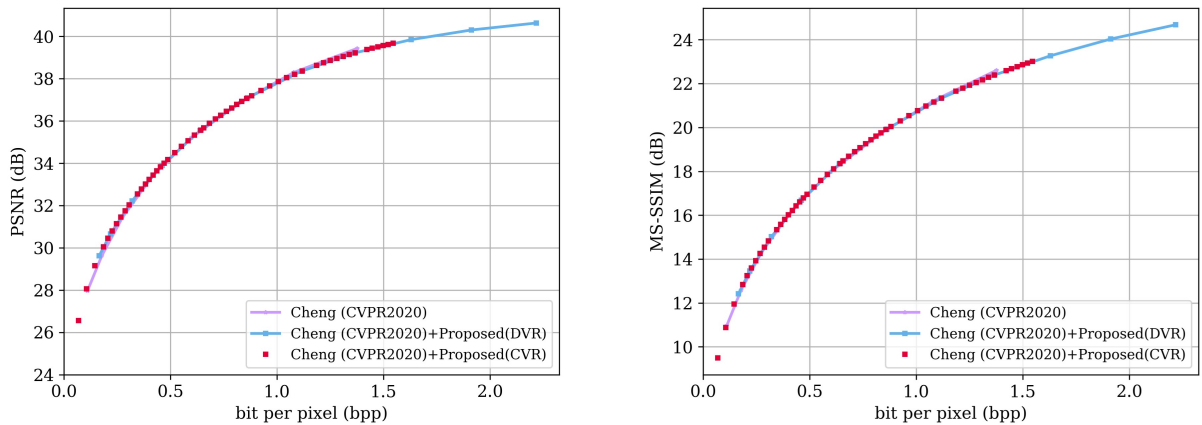

Fig. A2 Discrete/Continuous variable rate performance evaluation of QVRF using 8 fixed-rate models of Cheng *et al.* [8] as the benchmark.

**Table R1.** Results of bitrate consumption and compression performance on compressing kodim01 using the single model of QVRF with Cheng *et al.* [8] for discrete bitrates.

| Total bits (bpp) | Bits for latent representation (bpp) | Bits for side information (bpp) | PSNR (dB) | MS-SSIM |
|---|---|---|---|---|
| 0.24348 | 0.24112 | **0.00236** | 26.727 | 0.93857 |
| 0.30664 | 0.37215 | **0.00236** | 28.242 | 0.95894 |
| 0.55623 | 0.55387 | **0.00236** | 29.892 | 0.97330 |
| 0.79915 | 0.79679 | **0.00236** | 31.770 | 0.98339 |
| 1.0766 | 1.0743 | **0.00236** | 33.563 | 0.98944 |
| 1.3663 | 1.3640 | **0.00236** | 35.029 | 0.99280 |
| 1.6540 | 1.6516 | **0.00236** | 36.103 | 0.99467 |
| 1.9502 | 1.9479 | **0.00236** | 36.870 | 0.99576 |